# Magnetic field-induced enhancement of the nitrogen-vacancy fluorescence quantum yield

M. Capelli,[1] P. Reineck,[1] D. W. M. Lau,[1] A. Orth,[1] J. Jeske,[2] M. W. Doherty,[3] T. Ohshima,[4] A.D. Greentree[1] and B.C. Gibson[1]

[1] ARC Centre of Excellence for Nanoscale BioPhotonics, School of Science, RMIT University, Melbourne, VIC 3001, Australia

[2] School of Science, RMIT University, Melbourne, VIC 3001, Australia

[3] Laser Physics Centre, Research School of Physics and Engineering, Australian National University, Canberra, Australian Capital Territory 2601, Australia

[4] National Institutes for Quantum and Radiological Science and Technology, Takasaki, Gunma 370-1292, Japan

**The nitrogen-vacancy (NV) centre in diamond is a unique optical defect that is used in many applications today and methods to enhance its fluorescence brightness are highly sought after. We observed experimentally an enhancement of the NV quantum yield by up to 7% in bulk diamond caused by an external magnetic field relative to the field-free case. This observation is rationalised phenomenologically in terms of a magnetic field dependence of the NV excited state triplet-to-singlet transition rate. The theoretical model is in good qualitative agreement with the experimental results at low excitation intensities. Our results significantly contribute to our fundamental understanding of the photophysical properties of the NV defect in diamond and may enable novel NV centre-based magnetometry techniques.**

The negatively charged nitrogen-vacancy (NV) centre in diamond has attracted great interest as a solid state system for its range of physics, biological and chemical properties. Spin manipulation of the NV centre has been demonstrated for quantum computing and communication,[1,2] while its highly photostable emission at room temperature makes it a useful tool as a marker in biological applications[3–5] as well as a magnetic and electric field sensor.[6–9]

NV-based magnetometry is conventionally performed using optically detected magnetic resonance (ODMR). In ODMR experiments the NV centre emission is recorded as a function of the frequency of an applied microwave field. This frequency is swept around the zero-field resonant transition between the $m_s = 0$ and $m_s = \pm 1$ spin states of the NV centre at $D_{gs} \simeq 2.87$ GHz. The resonant transitions depend on the external magnetic field and can be detected optically because of the different emission intensities of the spin states.[10,11] Magnetic field sensing with an ensemble of NV centres in bulk diamond showed a sensitivity up to 0.9 pT/$\sqrt{\text{Hz}}$.[12] However, these experiments require a microwave source (e.g. a metallic wire connected to a microwave generator) to be within about 100 µm from the measurement position to deliver the microwave field efficiently. This can limit the use of ODMR in biology, where the integration of the microwave source into the biological system of interest can be challenging and in many cases impossible.

An alternative approach to NV-based magnetometry is to exploit the magnetic field dependence of the NV centre emission intensity in the absence of a microwave field.[13] Recently, accurate microwave-free magnetic field measurements have been demonstrated,[14] focusing on the NV centre emission change with precisely aligned high magnetic fields. Additionally, strongly focused light yielding high local excitation intensities of 220 mW (~400 MW/cm2) were used in these experiments.

Here we demonstrate experimentally that in the low optical excitation intensity regime (<100 W/cm2), far from NV saturation, the NV fluorescence is enhanced at magnetic field intensities above 100 mT by up to 7% with respect to the field-free case. We rationalise our findings by introducing a magnetic field-dependent transition phenomenologically from the excited NV triplet to the singlet state to the standard NV centre emission model. We find our model to be in good agreement with experimental results at the lowest excitation intensities investigated and in qualitative agreement at higher excitation intensities.

We used a custom-built confocal microscope with a 0.1 NA objective (see Fig. 1a), which focused the off-resonance 532 nm laser excitation (Laser quantum, gem 532) to a 0.99 ± 0.17 μm² focal spot (see ESI Fig. S1). The sample used in all experiments was a high-pressure high-temperature (HPHT) bulk diamond (Element Six) cut along the (001) crystallographic plane. It was electron-beam irradiated to a fluence of $10^{18}$ cm$^{-2}$ and subsequently annealed to obtain ~1 ppm NV density in the sample. We produced the magnetic field with a permanent magnet positioned at varying distances from the sample, keeping the magnetic field direction along the vertical axis in the laboratory frame as shown in Fig. 1b. The photoluminescence (PL) was filtered by a 532 nm notch filter and collected with an avalanche photodiode detector (APD, Excelitas, SPCM-AQRH-14). Emission spectra were collected with a SpectraPro spectrometer (Princeton Instruments) with a PIXIS CCD camera.

Fig. 2a shows the PL normalised to the zero-field value as a function of magnetic field for two excitation intensities of 0.2 μW and 70 μW (see ESI for measurement details). The initial decrease of PL with increasing magnetic field at < 50 mT has already been discussed in the literature.[14–18] Without an external magnetic field, the system is spin-polarized to the $m_s = 0$ spin state by the excitation light.[19] Introducing a magnetic field component not aligned along the NV symmetry axis causes the spin levels to mix, leading to a partial population of the $m_s = \pm 1$ spin states that increase with increasing magnetic field. From the excited $m_s = \pm 1$ spin states the system can transition non-radiatively to a metastable singlet state (see Fig. 1c), decreasing the overall emission. Fig. 2a shows a significant increase in emission intensity with an increasing magnetic field, even above the zero-field intensity in the regime of low excitation intensity, after reaching minima at 30 mT (0.2 μW) and 50

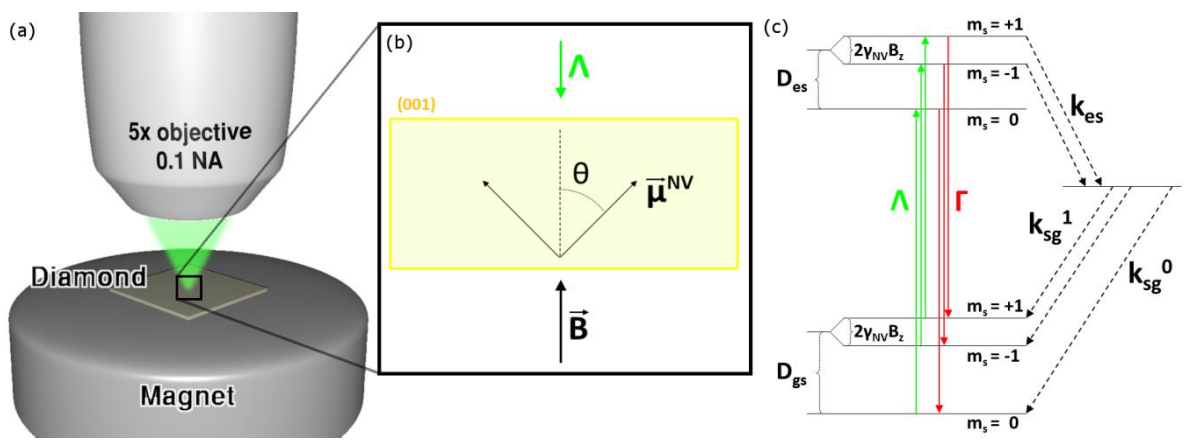

Fig. 1 (a) Schematic of the experimental setup. The 532 nm off-resonant excitation light was focused on the diamond sample by a 0.1 NA objective. The emission was then collected by the same objective for detection. We used a notch filter before the detector to remove the reflected excitation light. A permanent magnet produced the magnetic field needed in our experiments. (b) Representation of the relative directions between the external magnetic field B and the orbital magnetic moment of the NV centre, μ$^{NV}$. The orbital momentum is defined by the NV symmetry axis and depends on the crystallographic plane of the diamond sample, here: (001). The 532 nm laser's direction of incidence is indicated by the green arrow and its effective incoherent excitation rate by Λ. (c) Electron spin structure of the NV centre as implemented in our model. The parameters used were the effective incoherent excitation rate (Λ), the NV emission rate (Γ), the transition rate from the $m_s = \pm 1$ excited states to the metastable singlet state ($k_{es}$) and the transition rates from the singlet state back to the ground states ($k_{sg}^1$ and $k_{sg}^0$).

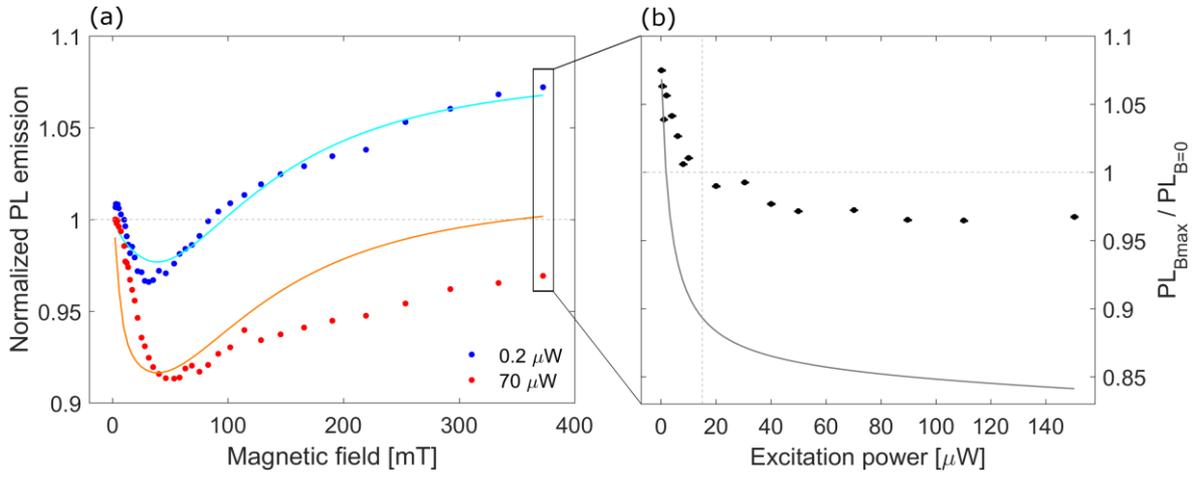

Fig. 2 (a) NV emission intensity as a function of external magnetic field for 70 µW (red) and 0.2 µW (blue) excitation power with a 532 nm laser. (b) Ratio of the PL intensity at $B_{max}$ = 372.6 mT and 0 mT as a function of excitation intensity. The emission intensity was measured at $B_{max}$ and then normalised to the respective zero-field value. The solid lines show the model fitted to the experimental data.

mT (70 µW). This has neither been reported to date to the best of our knowledge nor can it be explained with the currently accepted standard NV model.[20] Fig. 2b shows a monotonic decease of the ratio between the PL at a high magnetic field ($B_{max}$ = 372.6 mT) and the field-free case (B = 0 mT) as a function of excitation power. It is possible to identify a threshold value at around 15 µW where the PL intensity at $B_{max}$ and B=0 are identical. At lower excitation intensities the NV emission increases above the field-free intensity at high magnetic fields; at higher values it remains below the field-free intensity at high magnetic fields. At the highest excitation powers investigated the ratio decreases asymptotically towards a value of 0.96. A theoretical model that rationalizes the observed effects is explained in the following paragraph. It is in good qualitative agreement with the experimental results obtained at low excitation intensities (Fig. 2a, blue line), but deviates significantly at high excitation intensities (Fig. 2a, orange line and Fig. 2b, black line). Possible corrections for this discrepancy are investigated in ESI Fig. S2. Note that for the modelled results obtained at 70 µW excitation intensity (Fig. 2a, red dots) the fitting yields an excitation value one order of magnitude less than the one actually used in the experiments. We excluded ionization from NV⁻ to NV⁰ as a possible cause for any of the effects observed in Fig. 2 (see ESI Fig. S3 for details).

The standard NV model[20] consists of the seven-level system depicted in Fig. 1c. The modelling method we used as a baseline for this study is the same discussed in literature.[21] The ground state is a spin triplet system whose structure is described by the Hamiltonian

$$H_{gs} = D_{gs} \cdot S_z^2 + \gamma_{NV} \boldsymbol{B} \cdot \boldsymbol{S}, \qquad (1)$$

where $D_{gs} \simeq 2.87$ GHz is the zero-field splitting, $\boldsymbol{S} = (S_x, S_y, S_z)$ is the spin operator vector with the z-axis defined along the NV symmetry axis and $\gamma_{NV} = \frac{g_e \mu_B}{\hbar} \simeq 28$ GHz/T is the electron-spin gyromagnetic ratio. The excited state system has the same Hamiltonian as the ground state replacing $D_{gs}$ with the excited state zero-field splitting $D_{es} \simeq 1.4$ GHz.[22] From the excited state, the system can relax through a spin conserving radiative transition, with a rate $\Gamma = 1/\tau$ corresponding to a lifetime $\tau \simeq 12$ ns,[23] or through a non-radiative path involving the intersystem crossing to the singlet excited and ground states[24] modelled by us as a single level. The characteristic spin-dependent brightness of the NV centre originates from the coupling of the excited $m_s = \pm 1$ states with the metastable singlet state, but it is still not fully understood. We modelled the transition to the singlet state ($k_{es}$) as a combination of both the non-radiative transition rate to the excited singlet state ($\tau_{SING} = 24$ ns)[20] and the radiative singlet relaxation in the near infrared ($\tau_{IR} = 0.9$ ns):[25] $k_{es} = \frac{1}{\tau_{SING} + \tau_{IR}} = 0.04$ ns⁻¹.

The transition rates from the singlet ground state to the triplet ground state are taken from literature ($k_{sg}^0 = 1/440$ ns$^{-1}$ and $k_{sg}^1 = k_{sg}^0$).[15,25,26]

To model the dynamics of a quantum system such as the NV centre, we calculated the steady-state solution of the Lindblad master equation[27,28] implementing all the monodirectional transitions represented in Fig. 1c. Dephasing processes between spin states are considered for the triplet ground and excited states separately. According to this model, the emission of the NV centre decreases monotonically with increasing external magnetic field, and the quantum yield does not depend on the excitation intensity. Both predictions are not in agreement with the experimental findings reported here. To qualitatively reproduce our experimental results, two modifications are introduced to the standard NV model discussed thus far: 1. the spin-lattice relaxation time ($T_1$) is included and 2. a magnetic field dependence of the triplet-to-singlet transition rate $k_{es}$ is introduced. The former accounts for the experimentally observed excitation dependence, while the latter reproduces the emission increase at high magnetic fields. These processes are discussed in detail in the following. Only the implementation of both processes in our model predicts the observed significant increase of NV emission at high magnetic fields relative to B=0 (see Fig. 3, green line).

**1. Spin-lattice relaxation time ($T_1$):** In our regime of low optical excitation power, $T_1$ becomes relevant since it is on the same order of magnitude as the inverse of the excitation rate ($1/\Lambda \sim 0.18$ ms) and it counteracts the spin polarization effect of the excitation laser. Hence, including $T_1$ in the model leads to a less significant decrease of the NV emission with increasing magnetic field at low excitation power (see Fig. 3, red line). To implement $T_1$ in our model, all possible transitions between the three spin states ($m_s = 0, \pm 1$) are represented by the same transition rate of $k_{T_1} = 1/T_1$ (with $T_1 = 0.5$ ms)[29] in the excited and ground states separately. Since $T_1$ changes as a function of an applied magnetic field for either aligned[30] or misaligned[31] magnetic fields, we initially implemented this dependency in our model. However, this addition did not appear to impact the emission of the NV centre as a function of magnetic field (see ESI Fig. S4 for details), thus was removed to limit the number of free parameters in our model.

**2. Magnetic field dependent transition rate $k_{es}$:** We modelled the decay of the $k_{es}$ transition rate with magnetic field using the function

$$k_{es}(B) = \frac{k_{es}(0) - k_{es}(\infty)}{1 + B^2/B_0^2} + k_{es}(\infty). \qquad (2)$$

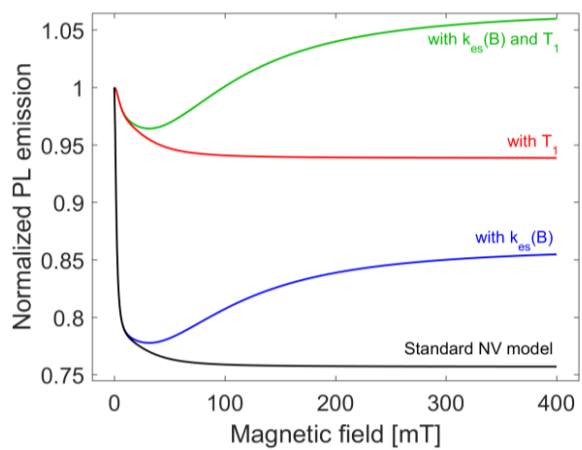

Fig. 3. Comparison between the NV emission intensity as a function of external magnetic field for the different steps implemented in the model.

The Lorentzian decay function resulted in the best fit to our data compared to other decay functions such as first order polynomial, exponential and Gaussian functions (see ESI Fig. S5). Using $k_{es}(\infty)$ and $B_0$ as free parameters we obtained $k_{es}(\infty) = 0.52 \pm 0.03$ and $B_0 = 130 \pm 10$ mT. This extension of the standard NV model qualitatively reproduces the observed minimum of the NV emission at about 50 mT and the increase in emission intensity for increasing magnetic fields as shown in Fig. 3, blue line.

A possible explanation for the decrease in the non-radiative transition rate ($k_{es}$) is a decrease in spin-orbit coupling between the triplet and the singlet state with increasing external magnetic field. The spin-orbit interaction could potentially change in the presence of a large misaligned field due to the reorientation of the spin quantization axis relative to the effective orbital magnetic moment. The latter is mixed by the electron orbitals and the structure of the defect, whereas the former is influenced by the applied magnetic field. The precise details of this mechanism are unclear and require further theoretical and experimental study.

The transition from the triplet excited state to the singlet state is the main non-radiative decay pathway for the excited NV centre. Hence, any change in this transition rate has a major effect on the NV emission intensity. At magnetic fields < 50 mT, the presence of a misaligned component of this magnetic field rapidly mixes the spin states, increasing the $m_s = \pm 1$ states population. This leads to a strong increase in the singlet state population, which dominates over any change in transition rate in particular at higher excitation intensities. At higher magnetic fields the spin states become almost equally populated, thus the transition rate $k_{es}$ decrease starts to affect the singlet state population significantly. Fig. 4 demonstrates these two competing effects: the mixing of spin states increasing and the transition rate $k_{es}$ decreasing with increasing external magnetic field. Together they affect the singlet state population, which mirrors the fluorescent emission of the NV centre shown in Fig. 2a. This model rationalizes both the decrease of NV emission at intermediate magnetic fields as well as the NV emission increase at higher magnetic field values.

We observe that the increase in NV emission at B$_{max}$ relative to B=0 is most pronounced (up to 7 % in our experiments) at low excitation intensities. As previously described, at low excitation intensities the spin polarisation of the NV centre is counteracted by the spin-lattice relaxation, which leads to significant population of the $m_s = \pm 1$ states, thus reducing the NV emission. Therefore, at low excitation intensities, the reduction in emission caused by the non-aligned external magnetic field is less significant, since the spin states are already mixed, and the change in transition rate $k_{es}$ proves to be the dominant effect in the overall emission with increasing magnetic fields (see Fig. 4a).

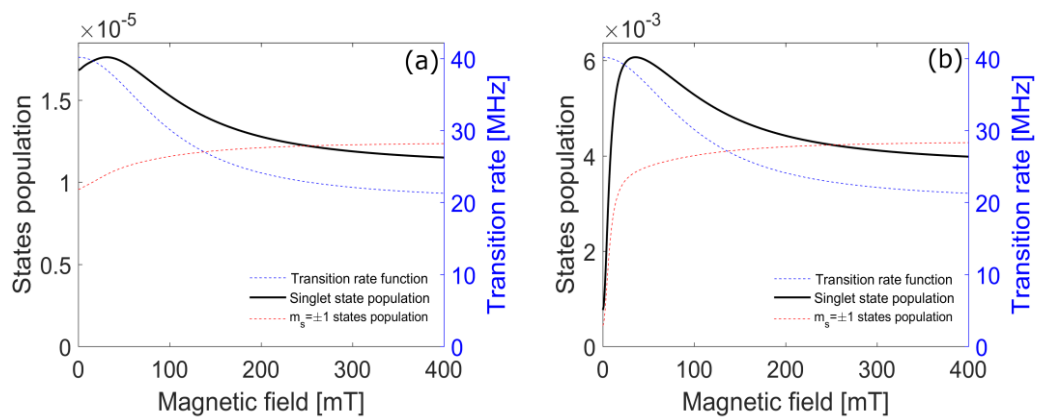

Fig. 4. Population of selected states and the transition rate function k$_{es}$ as a function of external magnetic field predicted by our theoretical model, for excitation intensities of 0.2 µW (a) and 70 µW (b). The singlet state population (solid black curve), the excited m$_s = \pm 1$ state population (dashed red curve) and the transition rate (dashed blue curve) are shown.

In conclusion, we have demonstrated that a misaligned external magnetic field can enhance the quantum yield of the NV centre emission at low excitation intensities. Such regime of low excitation intensity becomes particularly important in magnetometry for biological applications due to the need to limit the effects of phototoxicity. We rationalise our findings by introducing a magnetic field dependence to the triplet-to-singlet transition rate in the standard NV model. The model is in good qualitative agreement with experimental results at low excitation intensities but deviates significantly from experiments at high excitation intensities. Our findings are relevant for existing as well as the development of novel magnetometry approaches using the NV centre. Furthermore, they are a significant contribution towards a more complete theoretical understanding of the NV centre's photophysical properties.

## Acknowledgments

The authors would like to thanks Dr. Daniel Drumm and Prof. Jared Coles for the fruitful discussion on the theoretical modelling and the non-linear optimization. This work has been supported by ARC grants (FT110100225, LE140100131, CE140100003, DE170100169). M. C. acknowledges the support of RMIT University scholarship and B. C. G. acknowledges the support of an ARC Future Fellowship.

# Electronic Supplementary Information

**Focal spot size**

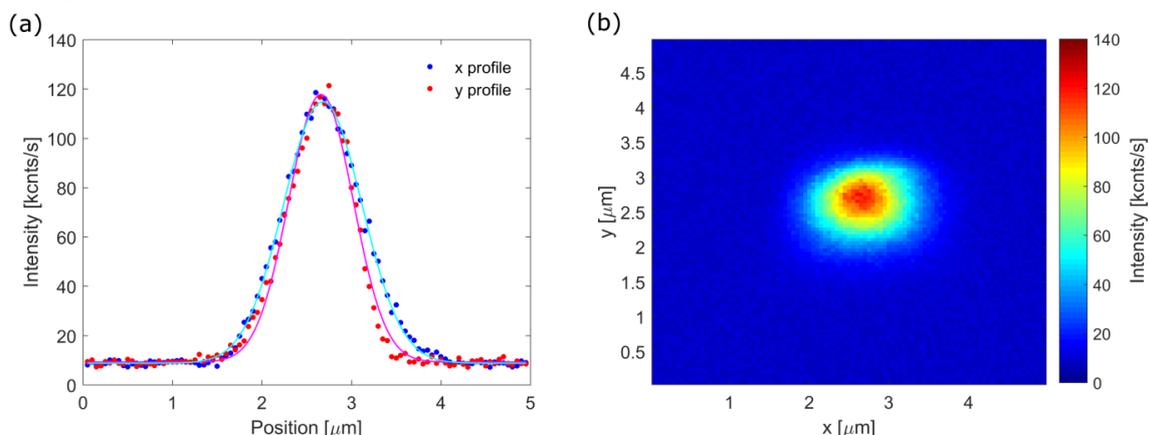

**Fig. S1** Fluorescent emission (a) profiles and (b) map for a 120 nm nanodiamond imaged with a 0.1 NA objective. The profiles are fitted with Gaussian function $A \cdot exp\left[-\frac{(x-x_0)^2}{\sigma_x^2}\right]$ where $\sigma_x$ represents the semi-axis along x (the same applies for y). The focal spot area has been calculated as $\pi\sigma_x\sigma_y$, and its error propagated from the standard deviation of the parameters obtained from the fitting algorithm.

**Experimental method and analysis**

Each data point in Fig. 2 represents the averaged value of a recorded fluorescence time-trace. Each time-trace lasted 60 s with an acquisition frequency of 20 samples/s. At the same time, the laser beam intensity was recorded with power meter (Thorlabs PM100D) software. We took into account the average laser intensity to correct small variations on excitation intensity during the experiment. Each data point was than normalized to the value recorded at 0 mT.

**High excitation power corrections to the NV centre emission**

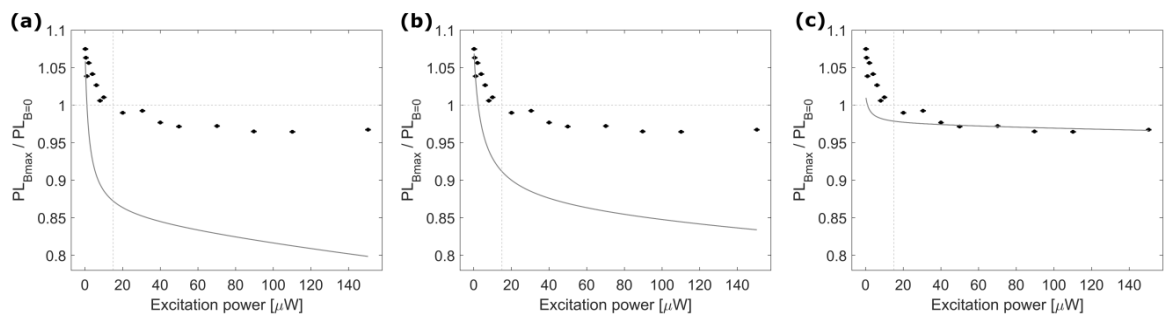

**Fig. S2** (a) Ratio of the PL intensity at $B_{max}$/B=0 mT as a function of excitation intensity as reported in Fig. 2b. We modelled different correction to try to reduce the discrepancy. (b) Fit of the data considering Gaussian-distributed excitation intensity. In the initial fit, the excitation rate used was constant over the entire focal spot. In this fit, a Monte Carlo simulation has been used to reproduce the difference in excitation intensity for multiple NV centres scattered at random distances from the centre of the focal spot. Applying such correction, the model better represents the experimental data, but the discrepancy at high excitation power is not completely solved. (c) Fit of the data considering a constant, magnetic field independent background fluorescence. Considering another emitter $E_0$ distributed with a density $n_{E0}$ over the sample and with a brightness $b$-times the brightness of an NV centre, the correcting factor was proportional to $\frac{n_{E_0}}{n_{NV}} \cdot b$. Applying such correction, the model better represents the experimental data at high excitation power but at the cost of creating a discrepancy at low excitation power. Moreover, to obtain the result shown in (c) a correction factor of 5 needs to be used, which we consider unlikely in our experiment for both the emitter density ($n_{E0}$) and brightness ($b$).

**Spectral analysis**

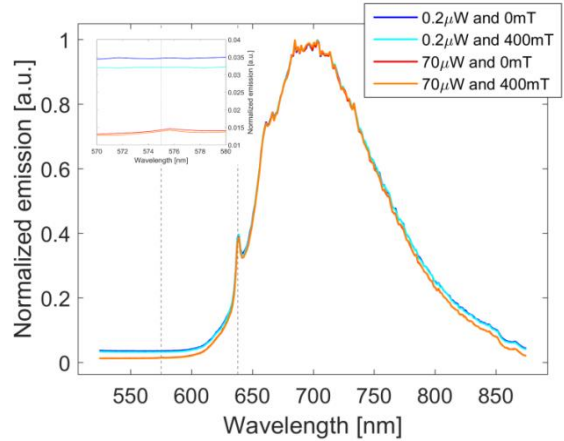

**Fig. S3** Spectra of ensembles of NV centres in HPHT single crystal diamond. The typical spectrum of the negatively charged NV centre (NV⁻) is shown, with the characteristic zero phonon line (ZPL) at 638 nm (right dashed line) and a broad phonon sideband emission with a maximum around 700 nm. Our sample does not show a measurable presence of the less fluorescent neutrally charged NV centre (NV⁰), whose ZPL should appear at 575 nm (left dashed line).[1] The four spectra shown correspond to the four combinations of maximum and minimum magnetic field and excitation power used in our experiments, filtered only by a notch filter around 532 nm. At a constant excitation power, the spectrum does not change with the magnetic field. The slight offset of the spectra at the different excitation power is caused by a greater integration time used to take the measurement at 0.2 µW which increased the background collected.

**Spin lattice relaxation time**

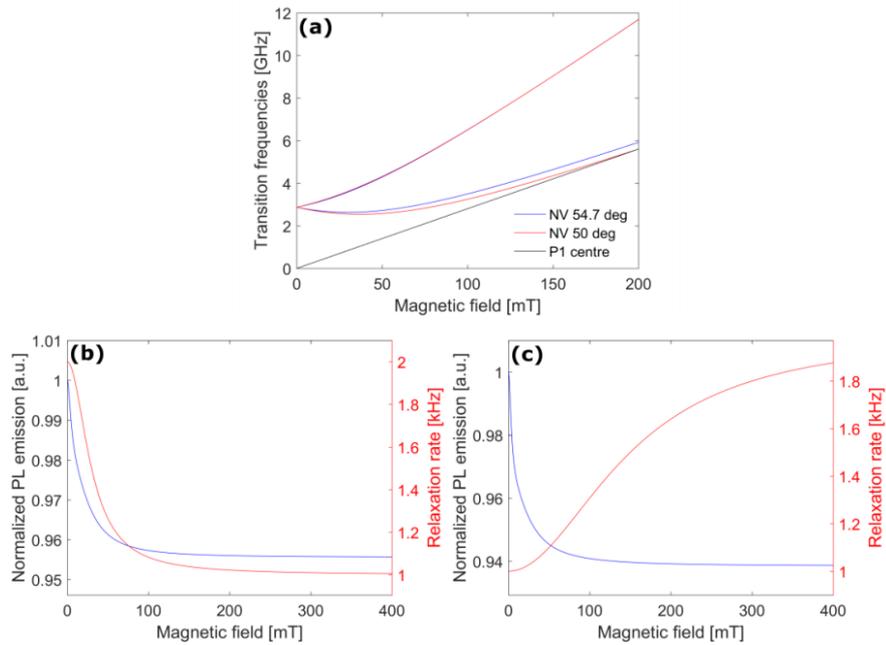

**Fig. S4** (a) Transition frequencies as a function of magnetic field for NV centres misaligned by 54.7 degrees with the magnetic field vector (blue curve), NV centres misaligned by 50 degrees (red curve), P1 centres (electron spin of a single nitrogen atom, black curve). (b, c) NV emission intensity as a function of external magnetic field with constant triplet-to-singlet rate ($k_{es}$ = 0.04 ns⁻¹) and changing spin relaxation rate ($k_{T_1}$) as a function of magnetic field. In both simulations, a low excitation rate (Λ = 5.4 kHz) is used. The simulation (b) shows the effect of a decreasing relaxation rate with increasing magnetic field. The relaxation rate decreases when an external magnetic field is applied to an ensemble of NV centre with different orientations because it removes the resonant transition at B=0 mT (a difference of 5 degree is enough to reduce the initial relaxation rate by half).[2] The simulation (c) shows the effect of an increasing relaxation rate with increasing magnetic field. The relaxation rate increases when the external magnetic field splits the electron spin energy levels of the surrounding nitrogen atoms (P1 centre), allowing a resonant transition between the NV centre and the P1 centre to occur (this resonance would happen at a specific magnetic field value if the NV centre were aligned to the magnetic field).[3]

Neither of the two changes in the spin relaxation rate have significant effect on the NV emission, even at low excitation rate.

## Rate functions comparison

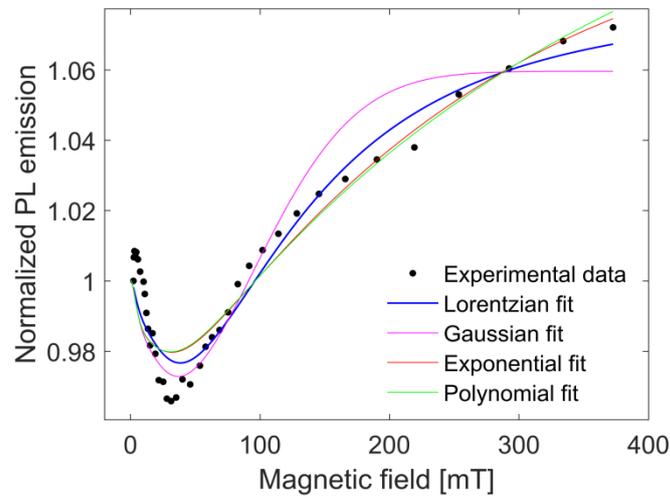

**Fig. S5** Black dots represent the same photoluminescence (PL) emission at 0.2 µW reported in Fig. 2a. The 4 curves show the fit of the model to the data, each of them using a different decaying function for the transition rate change with increasing magnetic field. Using the normalized residuals as a criterion, the Lorentzian fit emerges as the best representation of our data.